\documentclass[a4paper,10pt]{article}
\usepackage{type1cm,bm,dcolumn}
\usepackage{amsmath,amssymb,amsfonts}
\usepackage{graphics}
\usepackage[dvips]{graphicx,color}
\newcommand{\be}{\begin{equation}}
\newcommand{\ee}{\end{equation}}
\newcommand{\bea}{\begin{eqnarray}}
\newcommand{\eea}{\end{eqnarray}}
\newcommand{\ben}{\begin{enumerate}}
\newcommand{\een}{\end{enumerate}}
\newcommand{\bi}{\begin{itemize}}
\newcommand{\ei}{\end{itemize}}

\newcommand{\Ord}[2]{\mathcal O \left(#1\right)^{#2}}

\begin{document}
\title{Charged and rotating Black AdS Branes in $n+p+2$ dimensions}
\author{T. Delsate\footnote{terence.delsate(at)umons.ac.be}\\Theoretical and Mathematical Physics Dpt.,\\
Universit\'{e} de Mons - UMons, 20, Place du Parc, 7000 Mons - Belgium}
\date{\today}

\maketitle 

\begin{abstract}
We generalize the vacuum static black brane solutions of Einstein's equations with negative cosmological constant
recently discussed in literature, by introducing rotations and an electromagnetic field. We investigate numerically the thermodynamical properties of the charged and of the rotating $AdS$ black brane and we provide evidences for the existence of the charged and rotating case. In particular, we study the influence of the rotation and charge on the tension and mass. We find that the rotation essentially influences the tensions while the charge essentially influences the mass.
\end{abstract}

\section{Introduction}
The last decade has witnessed a growing interest in the physics of black holes in anti-de Sitter ($AdS$) backgrounds, motivated by the proposed correspondence relating $AdS$ spacetimes to conformal field theory defined on the boundary spacetime \cite{Maldacena:1997re,Witten:1998qj}, known as the $AdS/CFT$ correspondence. In this context asymptotically $AdS$ black hole solutions are related to thermal CFT. For example, the  Schwarzschild-AdS$_5$ Hawking-Page phase transition \cite{Hawking:1982dh} is interpreted as a thermal phase transition from a confining to a deconfining phase in the dual $D = 4$, ${\cal N} = 4$ super Yang-Mills theory \cite{Witten:1998zw}, while the phase structure of Reissner-Nordstr\"om-AdS (RNAdS) black holes, which resembles that of a van der Waals-Maxwell liquid-gas system is related to the physics of a class of field theories coupled to a background current \cite{Chamblin:1999tk}.

The construction of new asymptotically (locally) $AdS$ objects is then of particular interest in this context, since it provides new backgrounds for the dual CFTs. Black holes in $d$ dimensions have an horizon topology $S_{d-2}$, matching the topology of the conformal boundary. The simplest different topology (both at the horizon and at infinity) is $S_{d-3}\times S_1$ and has been first discussed in the $AdS/CFT$ context by \cite{Copsey:2006br} in $d=5$. Such objects are called black strings and have been constructed in arbitrary dimensions in \cite{rms}. Yet other types of objects have been proposed recently in \cite{adsbrane} where the horizon and asymptotic toplogy is $S_p\times \mathbb M_n$, $\mathbb M_n$ denoting a $n$-dimensional Ricci flat manifold. 

In this paper, we generalise the black brane type solutions constructed in \cite{adsbrane} to the charged and rotating case. In the section \ref{sec:model}, we present the model, the generic equations and boundary conditions and quickly review some known solutions. In section \ref{sec:asympt}, we present the asymptotic solution for large values of the radial coordinate and close to the horizon. Section \ref{sec:physics} is devoted to the thermodynamical and geometrical properties of the charged and/or rotating black brane. In particular we derive a Smarr law for the three cases (charged, rotating, charged-rotating). 
We present our numerical results in section \ref{sec:results}. This section is divided in three parts: charged case, rotating case and charged-rotating case. We discuss in more details the charged \emph{or} rotating cases and give evidences that the charged \emph{and} rotating case admits solutions. We finally draw our conclusions in the last section.

\section{The generic model, equations and boundary conditions}
\label{sec:model}
We consider the Maxwell-Einstein-Hilbert action in $d$ dimensions supplemented by a cosmological constant term:
\be
S = \frac{1}{16\pi G}\int_{\mathcal M} \sqrt{-g} \left( R - 2\Lambda - F^2\right) d^d x - \frac{1}{8\pi G}\int_{\partial\mathcal M}\sqrt{-h} K d^{d-1} x,
\label{smodel}
\ee
where $G$ is the $d$-dimensional Newton constant, $\mathcal M$ is the manifold under consideration, $\partial\mathcal{M}$ the boundary manifold $g$ is the determinent
of the metric, $R$ is the scalar curvature, $\Lambda$ is the cosmological constant,  $F=dA$ is the field strength of the Maxwell field $A$, $h$ is the determinent of
the boundary metric and $K$ is the extrinsic curvature of $\partial\mathcal M$ embedded in $\mathcal M$. Note that we use geometric units for the electromagnetic coupling. In the following, we will set the Newton constant to $G=1$.

We further define the AdS radius by $$ \ell^2 = -\frac{(d-1)(d-2)}{2\Lambda}.$$ In order to describe rotating brane like black objects, i.e. black objects with event horizon topology $S_p\times \mathbb M_n$, we introduce the following metric ansatz \begin{eqnarray}
& ds^2 = -b(r)dt^2 +a(r)d\vec z^2_n+  \frac{ dr^2}{f(r)} + g(r)\sum_{i=1}^{N-1}  \left(\prod_{j=0}^{i-1} \cos^2\theta_j \right) d\theta_i^2 \\  
&+h(r) \sum_{k=1}^N \left( \prod_{l=0}^{k-1} \cos^2 \theta_l  \right) \sin^2\theta_k \left( d\phi_k - w(r)  dt\right)^2 \nonumber \\ 
& +(g(r)-h(r)) \left\{ \sum_{k=1}^N \left( \prod_{l=0}^{k-1} \cos^2  \theta_l \right) \sin^2\theta_k  d\phi_k^2 \right. \nonumber\\
&\left.- \left[\sum_{k=1}^N \left( \prod_{l=0}^{k-1} \cos^2 \theta_l \right) \sin^2\theta_k   d\phi_k\right]^2 \right\},\nonumber
\label{metric-rot}
\end{eqnarray}
where $\theta_0 \equiv 0$, $\theta_i \in [0,\pi/2]$ for $i=1,\dots , N-1$, $\theta_N \equiv \pi/2$, 
$\phi_k \in [0,2\pi]$ for $k=1,\dots , N$, where $N=(p+1)/2$ and $d\vec z^2_n = \delta_{ij}dz^i dz^j,\ i,j=1,2\ldots,n$ denotes the transverse space. Here we take $p$ to be odd and $d=p+n+2$.

Note that the ansatz for the static black branes in arbitrary dimensions discussed in \cite{adsbrane} are recovered for $w(r)=0$, $h(r)=g(r)=r^2$.

The ansatz for the Maxwell field is given by
\be
A_a dx^a = V(r) dt+a_\varphi(r)\sum_{k=1}^N \prod_{l=0}^{k-1} \cos^2\theta_l \sin^2\theta_k d\phi_k.
\ee

The variation of the action \eqref{smodel} with respect to the metric components and the Maxwell field components leads to the Einstein-Maxwell equations given by
\be
G_a^b = -\Lambda \delta_a^b +  T_a^b,\  \nabla_a F^{ab} = 0,
\label{geneqs}
\ee
where $T_{ab} = F_{ac}F^c_{\ b} - \frac{1}{2} F_{cd}F^{cd}g_{ab}$ is the stress tensor and $G_a^b$ is the Einstein tensor. The independant components of $G_a^b$ and $T_a^b$ are given by
\bea
G_t^t&=&\frac{n f a''}{2 a}+\frac{1}{2} (p-1) \left(\frac{n f a' g'}{2 a g}+\frac{r^2 f w g' w'}{2 b g}+\frac{f' g'}{2 g}+\frac{f g''}{g}+\frac{f g'}{r g}+\frac{r^2}{g^2}\right)\\
&&+\frac{n r^2 f w a' w'}{4 a b}+\frac{n a' f'}{4 a}+\frac{n^2 fa'^2}{8 a^2}-\frac{3 n f a'^2}{8 a^2}+\frac{n f a'}{2 r a}-\frac{r^2 f w b' w'}{4 b^2}+\frac{r^2 w f' w'}{4 b}\nonumber\\
&&+\frac{r^2 f w w''}{2 b}+\frac{r^2 f w'^2}{4 b}+\frac{3 r f w w'}{2 b}+\frac{f'}{2 r}+\frac{(p-4) (p-1) f g'^2}{8 g^2}-\frac{p^2-1}{2g}\nonumber,
\eea
\bea
G_r^r&=&\frac{1}{2} (p-1) \left(\frac{n f a' g'}{2 a g}+\frac{f b' g'}{2 b g}+\frac{f g'}{r g}\right)+\frac{n f a' b'}{4 a b}+\frac{n^2 f a'^2}{8 a^2}-\frac{n f a'^2}{8 a^2}+\frac{n f a'}{2 r a}\nonumber\\
&&+\frac{f b'}{2 r b}+\frac{r^2 f w'^2}{4 b}+\frac{(p-2) (p-1) f g'^2}{8 g^2}-\frac{p^2-1}{2 g}+\frac{(p-1) r^2}{2 g^2},\nonumber
\eea
\bea
G_z^z &=&\frac{n f a''}{2 a}-\frac{f a''}{2 a}+\frac{n f a' b'}{4 a b}-\frac{f a' b'}{4 a b}+\frac{n a' f'}{4 a}-\frac{a' f'}{4 a}+\frac{n (p-1) f a' g'}{4 a g}\nonumber\\
&&-\frac{(p-1) f a' g'}{4 a g}+\frac{n^2 f a'^2}{8 a^2}-\frac{5 n f a'^2}{8 a^2}+\frac{n f a'}{2 r a}+\frac{f a'^2}{2 a^2}-\frac{f a'}{2 r a}\nonumber\\
&&+\frac{f b''}{2 b}+\frac{b' f'}{4 b}+\frac{(p-1) f b' g'}{4 b g}-\frac{f b'^2}{4 b^2}+\frac{f b'}{2 r b}-\frac{r^2 f w'^2}{4 b}\nonumber\\
&&+\frac{(p-3) f' g'}{4 g}+\frac{f' g'}{2 g}+\frac{f'}{2 r}+\frac{(p-1) f g''}{2 g}+\frac{(p-4)(p-1) f g'^2}{8 g^2}\nonumber\\
&&+\frac{(p-1) f g'}{2 r g}-\frac{p^2-1}{2 g}+\frac{(p-1) r^2}{2 g^2}\nonumber,
\eea
\bea
G_\theta^\theta &=&\frac{n f a''}{2 a}+\frac{n f a' b'}{4 a b}+\frac{n a' f'}{4 a}+\frac{n (p-2) f a' g'}{4 a g}+\frac{(n-3) n f a'^2}{8 a^2}+\frac{n f a'}{2 r a}\nonumber\\
&&+\frac{f b''}{2 b}+\frac{b' f'}{4 b}+\frac{(p-2) f b' g'}{4 b g}-\frac{f b'^2}{4 b^2}+\frac{f b'}{2 r b}-\frac{r^2 f w'^2}{4 b}+\frac{(p-2) f' g'}{4 g}\nonumber\\
&&+\frac{f'}{2 r}+\frac{(p-2) f g''}{2 g}+\frac{(p-5) (p-2) f g'^2}{8 g^2}+\frac{(p-2) f g'}{2 r g}+\frac{(p-5) r^2}{2 g^2}\nonumber\\
&&-\frac{(p-3) (p+1)}{2 g}\nonumber,
\eea
\bea
G_{\varphi_1}^{\varphi_2}&=&-\frac{n r^2 f w a' w'}{4 a b}+\frac{n f a' g'}{4 a g}-\frac{n f a'}{2 r a}+\frac{f b' g'}{4 b g}+\frac{r^2 f w b' w'}{4 b^2}-\frac{f b'}{2 r b}\nonumber\\
&&-\frac{r^2 w f' w'}{4 b}-\frac{(p-1) r^2 f w g' w'}{4 b g}-\frac{r^2 f w w''}{2 b}-\frac{r^2 f w'^2}{2 b}-\frac{3 r f w w'}{2 b}+\frac{f' g'}{4 g}\nonumber\\
&&-\frac{f'}{2 r}+\frac{f g''}{2 g}\nonumber,
\eea
\bea
G_t^\varphi &=& -\frac{n f w a' b'}{4 a b}+\frac{(n-1) r^2 f w^2 a' w'}{4 a b}+\frac{r^2 f w^2 a' w'}{4 a b}+\frac{n f a' w'}{4 a}+\frac{n f w a'}{2 r a}\nonumber\\
&&-\frac{f w b''}{2 b}-\frac{w b' f'}{4 b}-\frac{(p-1) f w b' g'}{4 b g}-\frac{r^2 f w^2 b' w'}{4 b^2}-\frac{f b' w'}{4 b}+\frac{f w b'^2}{4 b^2}\nonumber\\
&&+\frac{r^2 w^2 f' w'}{4 b}+\frac{(p-1) r^2 f w^2 g' w'}{4 b g}+\frac{r^2 f w^2 w''}{2 b}+\frac{r^2 f w w'^2}{b}+\frac{3 r f w^2 w'}{2 b}+\frac{1}{4} f' w'\nonumber\\
&&+\frac{w f'}{2 r}+\frac{(p-1) f g' w'}{4 g}+\frac{(p-1) f w g'}{2 r g}+\frac{1}{2} f w''+\frac{3 f w'}{2 r}-\frac{(p-1) r^2 w}{g^2},\nonumber
\eea

\bea
T_t^t&=&\frac{f w^2 a_\varphi'^2}{2 b}-\frac{f a_\varphi'^2}{2 r^2}-\frac{(p-1) a_\varphi^2}{g^2}-\frac{f V'^2}{2 b}\nonumber,\\
T_r^r&=&-\frac{f w a_\varphi' V'}{b}-\frac{f w^2 a_\varphi'^2}{2 b}+\frac{f a_\varphi'^2}{2
   r^2}-\frac{(p-1) a_\varphi^2}{g^2}-\frac{f V'^2}{2 b}\nonumber,\\
T_z^z&=&\frac{f w a_\varphi' V'}{b}+\frac{f w^2 a_\varphi'^2}{2 b}-\frac{f a_\varphi'^2}{2
   r^2}-\frac{(p-1) a_\varphi^2}{g^2}+\frac{f V'^2}{2 b}\nonumber,\\
T_\theta^\theta&=&\frac{f w a_\varphi' V'}{b}+\frac{f w^2 a_\varphi'^2}{2 b}-\frac{f a_\varphi'^2}{2
   r^2}-\frac{(p-5) a_\varphi^2}{g^2}+\frac{f V'^2}{2 b}\nonumber,\\
T_{\varphi_1}^{\varphi_2}&=&\frac{f w a_\varphi' V'}{b}+\frac{f w^2 a_\varphi'^2}{b}-\frac{f a_\varphi'^2}{r^2}+\frac{4a_\varphi^2}{g^2}\nonumber,\\
T_t^\varphi&=&-\frac{f w^2 a_\varphi' V'}{b}+\frac{f a_\varphi' V'}{r^2}-\frac{f w V'^2}{b}\nonumber.
\eea
The two independant Maxwell equations read
\bea
&&-\frac{n f w a' a_\varphi'}{2 a b}-\frac{n f a' V'}{2 a b}-\frac{f w a_\varphi''}{b}+\frac{f w a_\varphi' b'}{2 b^2}-\frac{w a_\varphi' f'}{2 b}-\frac{(p-1) f w a_\varphi' g'}{2 b g}\\
&&-\frac{f a_\varphi' w'}{b}-\frac{f w a_\varphi'}{rb}+\frac{f b' V'}{2 b^2}-\frac{f' V'}{2 b}-\frac{(p-1) f g' V'}{2 b g}-\frac{fV''}{b}-\frac{f V'}{r b}=0\nonumber,\\
 \nonumber\\
&&-\frac{n f w^2 a' a_\varphi'}{2 a b}+\frac{n f a' a_\varphi'}{2 r^2 a}-\frac{n f w a'V'}{2 a b}-\frac{f w^2 a_\varphi''}{b}+\frac{f a_\varphi''}{r^2}+\frac{f a_\varphi'b'}{2 r^2 b}\nonumber\\
&&+\frac{f w^2 a_\varphi' b'}{2 b^2}-\frac{w^2 a_\varphi' f'}{2 b}-\frac{(p-1) f w^2 a_\varphi' g'}{2 b g}-\frac{2 f w a_\varphi' w'}{b}-\frac{f w^2 a_\varphi'}{r b}+\frac{a_\varphi' f'}{2 r^2}\nonumber\\
&&+\frac{(p-1) f a_\varphi' g'}{2 r^2 g}-\frac{f a_\varphi'}{r^3}-\frac{2 (p-1) a_\varphi}{g^2}+\frac{f w b' V'}{2 b^2}-\frac{w f' V'}{2 b}-\frac{(p-1) f w g' V'}{2 b g}\nonumber\\
&&-\frac{f w V''}{b}-\frac{f V' w'}{b}-\frac{f w V'}{r b}=0\nonumber.
\eea

It is possible to express the equations as a system of 7 non linear coupled ordinary differential equations for the functions $a,b,f,g,w,V,a_\varphi$ by combining suitably the Einstein equation. However the resulting equations are very long and we refrain to write them here.

It should be stressed that it is possible to rewrite the equation for $w$ (resp. $V$) as a total derivative in the neutral case, where $V=a_\varphi=0$ (resp. in the non rotating case where $w=a_\varphi=0, g=h=r^2$):
\be
\left(  w'r^3\sqrt{ g^{p-1}a^n \frac{f}{b} }\right)'=0,
\ee
for the neutral rotating case and
\be
\left(  V'r^p\sqrt{ a^n \frac{f}{b} }\right)'=0,
\ee
for the charged non rotating case.
This leads to first integrals (say $j$ and $q$) associated respectively to the angular momentum and to the charge. Note that the function $a_\varphi$ does not appear in these particular case since it is associated with the magnetic field which is excited by the rotation in the charged case.

In order to have a well posed problem, we need $13$ boundary conditions, given by
\bea
&&f(r_h)=0,\ b(r_h)=0,\ b'(r_h)=b_1,\ a(r_h)=a_h,\ V(r_h)=V_h,\ w(r_h)=w_h,\nonumber\\
&&\Gamma_1(r_h)=0,\ \Gamma_2(r_h)=0,\ \Gamma_3(r_h)=0\\
&&g\rightarrow r^2,\ w\rightarrow0,\ V\rightarrow0,\ a_\varphi\rightarrow0\mbox{ for }r\rightarrow\infty,\nonumber
\eea
where $a_h,b_1$ are real constants adjusted in order to find asymptotically locally $AdS$ solutions and for some constants $V_h,\ w_h$ and where $\Gamma_i$ are combinations of the various functions, which should vanish at the horizon for the equations to be regular.
The $\Gamma_i$ are given by
\bea
\Gamma_1(r)&=& b' r \left(g (-f' g'+2 p+2)-8 a_\varphi^2\right)+r^3 \left(f'g^2 w'^2-2 (p+1) b' \right)+2 b' f' g^2\nonumber\\
\Gamma_2(r)&=& \frac{ar^2 \left(f' g^2 w'^2-2 b' (p-1)\right)}{b' f' g^2}+\frac{2 a}{r}-a'\\
\Gamma_3(r)&=& r^2 \left(2 a_\varphi b' (p-1)+f' g^2 w' (a_\varphi'w +V'\right)-a_\varphi' b' f' g^2\nonumber
\eea

\subsection{Known solutions}
Atlhough it seems hopeless to find an analytic solution to the full set of equations \eqref{geneqs}, there are some particular cases where an analytic solution exists.

For instance for $p=0$ or $n=0$, the metric describes a black hole (resp. topological black hole) and the vacuum non-rotating case is given by
\be
f(r)=\frac{r^2}{\ell^2}+1-\left(\frac{r_0}{r}\right)^{p},\ b(r)=f(r),\ g(r)=h(r)=r^2,\ \omega(r)=0,
\ee
for $n=0$ and 
\be
f(r)=\frac{r^2}{\ell^2}-\left(\frac{r_0}{r}\right)^{p},\ b(r)=f(r),\ a(r)=r^2,
\ee
for $p=0$.

In the same limits, the electrovacuum solution have been constructed in \cite{Gibbons:2004uw}.
The charged and rotating black hole has been constructed numerically in \cite{kunz}.
The case $n=1$ has been studied in \cite{brs} for the charged or rotating case and in \cite{rms} for the non rotating case. 

For a vanishing cosmological constant, the function $a$ can be set to $a(r)=1$ since the extradirections are Ricci-flat and the remaining $p+2$ dimensional space can be taken to be the charged and/or rotating $(p+2)$-dimensional black hole \cite{mp,bhchrot} (see \cite{bhrev} and references therein for a review). 

The phase structure of the case $n=1$ has been intensively studied \cite{gl,gubser,wiseman,rms,adsstab} (see \cite{bsrev} and reference therein for a review) with or without a cosmological constant.

\section{Near Horizon and asymptotic expansion}
\label{sec:asympt}
Far from the event horizon, the functions $a,b,f,g,w,V,a_\varphi$ obey the following expansion:
\begin{eqnarray}
\nonumber 
a(r)&=&\frac{r^2}{\ell^2}+\sum_{j=0}^{\lfloor(d-4)/2\rfloor}a_j(\frac{\ell}{r})^{2j}
 + \sum_k\delta_{d,2k+1}\zeta\log(\frac {r}{\ell}) (\frac{\ell}{r})^{d-3}
+c_z(\frac{\ell}{r})^{d-3}+\Ord{\frac{1}{r}}{d-1},
\\
\label{odd-inf}
b(r)&=&\frac{r^2}{\ell^2}+\sum_{j=0}^{\lfloor(d-4)/2\rfloor}a_j(\frac{\ell}{r})^{2j}
 + \sum_k\delta_{d,2k+1}\zeta\log (\frac {r}{\ell}) (\frac{\ell}{r})^{d-3}
+c_t(\frac{\ell}{r})^{d-3}+\Ord{\frac{1}{r}}{d-1},\nonumber
\\
f(r)&=&\frac{r^2}{\ell^2}+\sum_{j=0}^{\lfloor(d-4)/2\rfloor}f_j(\frac{\ell}{r})^{2j}
 + \sum_k\delta_{d,2k+1}(n+1)\zeta\log (\frac {r}{\ell}) (\frac{\ell}{r})^{d-3}\nonumber\\
&&+(n c_z+c_t + (p-1)c_g +c_0)(\frac{\ell}{r})^{d-3}+\Ord{\frac{1}{r}}{d-1},\nonumber\\
\nonumber
g(r)&=& r^2\left( 1 + c_g\left(\frac{\ell}{r}\right)^{d-5} \right)+\Ord{\frac{\ell}{r}}{d-1}\\
\nonumber
w(r)&=& c_w\left(\frac{\ell}{r}\right)^{d-3} + \Ord{\frac{\ell}{r}}{d-1}\\
\nonumber
V(r_)&=& c_V\left(\frac{\ell}{r}\right)^{d-3} + \Ord{\frac{\ell}{r}}{d-1}\\
\nonumber
a_\varphi(r)&=& c_\varphi\left(\frac{\ell}{r}\right)^{d-3} + \Ord{\frac{\ell}{r}}{d-1}
\end{eqnarray}   
where $d=n+p+2$, the constants $a_i,f_i,\zeta$ depend on $p,n,\ell$ and are given by $a_0= \frac{p-1}{n+p-1}$, $a_1=\frac{n (p-1)^2}{(n+p-3) (n+p-1)^2 (n+p)}$, 
$a_2=-\frac{n (p-1)^3 \left(n^2+n (4 p-17)+3 (p-3) p\right)}{3 (n+p-5) (n+p-3) (n+p-1)^3 (n+p)^2} $, $f_0=\frac{(p-1) (2 n+p)}{(n+p-1) (n+p)}$, 
$f_1=\frac{n (n+1) (p-1)^2}{(n+p-3) (n+p-1)^2 (n+p)}$, $f_2= -\frac{n (n+1) (p-1)^3 (n (p-4)+(p-3) p)}{(n+p-5) (n+p-3) (n+p-1)^3 (n+p)^2}$, 
$\zeta = -1/12$ for $p=2,n=1$, $\zeta=-3/400$ for $p=2,n=3$, $-1/100$ for $p=3,n=2$,$\ldots$, $c_0= 0$ for $p=2,n=1$, $c_0=3/800$ for $p=3,n=2$, $1/100$ 
for $p=3,n=2$,$\ldots$ and where $\lfloor X \rfloor$ denotes the floor integer value of $X$. The coefficient $c_t,c_z,c_g,c_w,c_\varphi,c_V$ are to be 
determined numerically and are related to the mass, tension, angular momentum, charge and magnetic momentum of the $AdS$ brane. Note that the constants 
$a_i,f_i,\zeta$ are the same than in the uncharged case \cite{adsbrane}.
Note also that the equations are invariant under arbitrary scaling of the metric functions $a,b$, related to the freedom of redifining the $t$ and $z$ coordinates.
We fix the normalisation of these two functions by adjusting the real constants $a_h,b_1$ described in the previous section such that the functions $a,b\rightarrow r^2/\ell^2$ at large values of $r$.

Close to the horizon, the metric and maxwell functions obey the following expansion
\bea
a(r)&=&a_h + a_1(r-r_h)+\Ord{r-r_h}{2},\nonumber\\
b(r)&=&b_1(r-r_h)+\Ord{r-r_h}{2},\nonumber\\
f(r)&=& f_1(r-r_h)+\Ord{r-r_h}{2},\nonumber\\
g(r)&=&g_h + g_1(r-r_h)+\Ord{r-r_h}{2},\\
w(r)&=&w_h + w_1(r-r_h)+\Ord{r-r_h}{2},\nonumber\\
V(r)&=&V_h + V_1(r-r_h)+\Ord{r-r_h}{2},\nonumber\\
a_\varphi(r)&=&A_h + A_1(r-r_h)+\Ord{r-r_h}{2},\nonumber\\
\eea
where the various coefficients can be expressed as functions of $a_h,b_1,g_h,w_h,V_h$.

The case of no rotation is much simpler: the functions $w,a_\varphi$ vanish, $g=h=r^2$ and the near horizon series reduces to
\bea
f(r)&=&\left(-\frac{q^2 r_h a_h^{-n} r_h^{-2p}}{n+p}+\frac{r_h}{(n+p+1)\ell^2}+\frac{p-1}{r_h}\right)(r-r_h)+\Ord{r-r_h}{2},\nonumber\\
a(r)&=&a_h + \frac{2 a_h r_h \left(q^2\ell^2-a_h^n (n+p) (n+p+1)  r_h^{2p}\right)}{q^2 r_h^2\ell^2-a_h^n (n+p) r_h^{2p}}(r-r_h) + \Ord{r-r_h}{2}\nonumber\\
b(r)&=& b_1(r-r_h) + \Ord{r-r_h}{2}
\eea

The requirement that $f'(r_h)>0$ leads to the following condition on the charge:
\be
q^2\leq\frac{a_h^n (n+p) r_h^{2 p-2} \left(\ell^2 (p-1)+r_h^2 (n+p+1)\right)}{\ell^2},
\label{ext_bound}
\ee
this condition is in agreement with the bound found in ref \cite{brs} for $n=1$. If the bound is saturated, $r_h$ becomes a double root of $f$ and the temperature vanishes; this is the extremal solution. The lower horizon radius should lead to a lower bound on the mass for a given value of the charge. We will discuss this further in the next sections.

Note that the maximal value of $q$ is a growing function of $r_h$ i.e. for a fixed value of the charge, we have a lower bound on the horizon radius or equivalently a maximal bound on the $AdS$ length.

\section{Physical properties of the $AdS$ branes}
\label{sec:physics}
\subsection{Thermodynamical properties}
The metric admits $\xi=\partial_t + \Omega_H^{(i)} \partial_{\varphi_i}$ as a killing vector. The $\Omega_H^{(i)}$s are fixed by the requirement that $\xi$ 
is null on the event horizon, leading to
\be
\Omega_H^{(i)}= \omega_h.
\ee

The asymptotical thermodynamical quantities are computed using the counterterm procedure described in \cite{counter} and applied to $AdS$ branes in \cite{adsbrane}, leading to the mass $M$, tension $\mathcal T$ and angular momentum $J$:
\bea
M &=& \frac{\ell^{p-1} V_p \mathcal V_{n}}{16\pi}((n+p)c_t - n c_z + (p-1)c_g) +M_c,\\
\mathcal T_i &=& \frac{\ell^{p-1} V_p\mathcal V_{n}}{16\pi n L_i}(c_t - (p+1) c_z-(p-1)c_g) +\frac{1}{n} \mathcal T_c,\nonumber\\
J&=&\frac{\ell^{p-1} V_p \mathcal V_{n}}{16\pi}\frac{2(n+p+1)}{(p+1)}c_w,\nonumber
\eea
where $V_p$ denotes the surface of the unit $p$-sphere, $\mathcal V_n$ is the volume of the $n$ extradirections, $L_i$ is the length of the $i-th$ extradirection 
and where $M_c = -\mathcal T_c=$ are Casimir like terms appearing in odd dimensions (see ref. \cite{adsbrane}) and are not useful for our purpose.
Note that we add manually a $1/n$ factor in the definition of the tension. First, since there are $n$ extradirections, all playing a spectator role in the equations, all the tensions should be the same. Second, as we will see later, each extradirection will contribute one time in the Smarr relation if we use this definition.

For a charged non-rotating solution, the electric field with respect to a constant $r$ hypersurface is given by $E^{\mu}=g^{\mu\rho}F_{\rho\nu}n^{\nu}$, $n^\mu$ being the unit normal to the constant $r$ surface. The electric charge of the charged solutions is computed using Gauss' 
law by evaluating the flux of the electric field at infinity:
\begin{eqnarray} 
\label{Qc}  
Q=\frac{1}{4\pi G_d}\oint_{\Sigma }d^{d-2}y\sqrt{\sigma}u^{a}n ^{b}F_{ab}.
\end{eqnarray} 
If $A_{\mu}$ is the electromagnetic potential, then the electric 
potential $\Phi$, measured at infinity with respect to the horizon 
is defined as \cite{potential}:
\be
\Phi=A_{\mu}\chi^{\mu}|_{r\rightarrow\infty}-A_{\mu}\chi^{\mu}|_{r=r_h}~,
\ee
with $\chi^{\mu}$ a  Killing vector orthogonal to and null on the horizon.

In the case of rotating solutions, the charge is defined in the same way but the magnetic field is excited by the rotation and leads to a magnetic momentum. The electric charge and magnetic momentum are given by
\be
Q=\frac{\ell^{p-1}V_p\mathcal V_{n}}{16\pi} (n + p-1)c_V,\ \mu =\frac{\ell^{p-1}V_p\mathcal V_{n}}{16\pi} (n + p-1)c_\varphi.
\ee

The temperature and entropy are computed in the standard way, leading to
\be
T_H = \frac{\sqrt{b_1 f_1}}{4\pi},\ S = \frac{V_p \mathcal V_{n}}{4}\sqrt{a_0^ng_h^{p-1}}r_h ,
\ee
where $f_1$ denotes the derivative of $f$ at the horizon.

We derived a Smarr law for the rotating $AdS$ brane and for the charged brane using the same approach than in ref. \cite{brs}:
\bea
M +  \mathcal T_i L_i = T_H S +  \frac{(p+1)}{2}\omega_h J,\\
M +  \mathcal T_i L_i = T_H S +  \Phi Q,\nonumber
\eea
where summation over the repeated index is understood.

Although it seems reasonable that the Smarr law in the charged-rotating case should read $M + \mathcal T_i L_i = T_H S +  \frac{(p-1)}{2}\omega_h J + \Phi Q$, we could not proove this using the standard technique. This is due to the fact that in the purely charged or purely rotating case, the charge (resp. the angular momentum) appear as a first integral of the equations; in the charged \emph{and} rotating case, they are not constant anymore but rather asymptotic constant. The Smarr law we derived reads
\be
M + \mathcal T_i L_i = T_H S +  \frac{(p+1)}{2}\omega_h J_H,
\ee
where $J_H$ is the angular momentum of the event horizon defined by
\be
J_H = \frac{V_p}{16\pi}\frac{2}{p-1}\sqrt{\frac{f_1}{b_1}g_h^{p-1}a_h^n}r_h^3 w_1.
\ee
This can be understood by the fact that the electromagnetic field carries a part of the total angular momentum.

\subsection{Geometric properties properties of the $AdS$ brane}
The equations admit another scaling property:
\bea
&&r\rightarrow \lambda r,\ \ell\rightarrow\lambda \ell,\ M\rightarrow,\lambda^{p-1}M\ \mathcal T\rightarrow\lambda^{p-1}\mathcal T,\ w_h\rightarrow\lambda^{-1} w_h,\ J\rightarrow\lambda^{p} J,\nonumber\\
&& Q\rightarrow\lambda^{p-1}Q,\ \mu\rightarrow\lambda^{p-1}\mu,\ T_H\rightarrow\lambda^{-1}T_H,\ S\rightarrow\lambda^pS.
\label{scaler}
\eea

The horizon topology is $\mathbb M_n\times S_p\times\mathbb R$, $\mathbb M_n=\prod_{i=1}^n L_i$ being the $n$-dimensional manifold describing the extradirections. The setup can be interpreted in different ways (such as a $n$ dimensional infinite brane surrounded by a $(p+1)$-dimensional black hole, a $p+2$ dimensional black hole with a $n$ dimensional compact internal space, etc \cite{adsbrane} depending on the range of the $L_i$ (wich can be infinite). The geometry can be a sort of $AdS$ with compact directions, which can be seen as the global $AdS$ with indentifications, this is however not clear from a more formal point of view and regarding the asymptotic spacetime.

\section{Numerical Results}
\label{sec:results}
In this section, we present our numerical results and discuss the thermodynamics of the charged and of the rotating case. Then, we present a solution to the charged \emph{and} rotating case, but don't study systematically its properties, due to some restrictions in our approach, that will be discussed further in the following.

It should be stressed that in the case where the rotation is absent, the metric ansatz reduces to the metric ansatz considered in \cite{adsbrane}. 
In this case, we don't have restrictions on the value of $p$.

\subsection{Charged case}
We integrated numerically the system of equations \eqref{geneqs} supplemented by the additional requirement that $g=h=r^2, w=a_\varphi=0$, using the solver Colsys \cite{colsys} for various values of $p,n,\ell,q,r_h$. We adopted the following numerical approach: we first fix the value of $a,b,V$ at the horizon and then rescale the solution in order to follow the suitable asymptotic behaviour. We typically vary the cosmological constant and use the scaling properties \eqref{scaler}
in order to fix the $AdS$ radius to $\ell=1$. This approach suffers from a considerable drawback, namely the fact that it is complicated to obtain a fixed value of the charge or of the electric potential at the origin; one has to produce a large amount of data in order to foliate the parameter space and find constant charge (or other thermodynamical quantities) slices. However, in certain regions of the parameters, some physical quantities such as the electric potential at the horizon do not vary too much and it is possible to have ideas of the underlying physics at fixed eletric potential. This is the reason why we present the physical quantities for $n=2,\ p=3$ only, hoping that it catches the main features of the generic solutions. Note that we cosntructed the solutions for others values of $p,\ n$ but we did not study systematically the thermodynamics.
Note also that the case $n=1$ has already been considered in \cite{brs}.

\begin{figure}
 \centering
  \includegraphics[scale=.5]{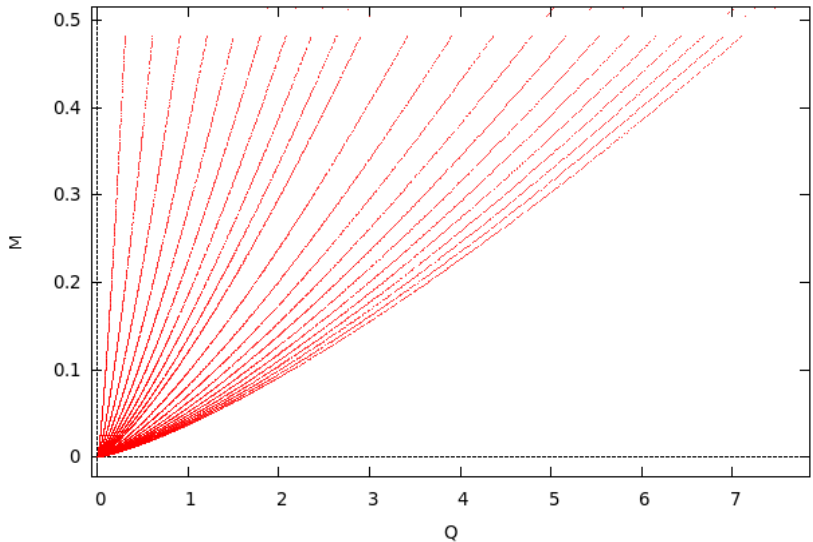}
\caption{Domain of the solution in the $Q-M$ plane for $p=3,\ n=2$ and $\ell=1$. Here we plot the absolute value of the charge. }
\label{fig:n2p3QM}
\end{figure}

First, we found numerical evidences that for given values of the charge, there exists a lower bound on the mass, along with the discussion following equation \eqref{ext_bound}. This is illustrated on figure \ref{fig:n2p3QM} where we plot the values of the mass versus the values of the charge for a large number of solutions. The approach was the following: we varied the horizon radius for each gradually increasing value of the electric potential at the horizon. The mass seems to approach a lower bound as can be appreciated on the figure.

As the mass approaches the lower bound, a critical solution is approached and the numerical solver fails to converge.
The study of this critical solution is technically involved and seems to require a different parametrization of the metric ansatz. This is beyond the purposes of this paper.

The existence of a minimal mass for fixed values of the charge is to be interpretted as the balance between repulsive
electrostatic and attractive gravitational interactions.

\begin{figure}
 \centering
  \includegraphics[scale=.5]{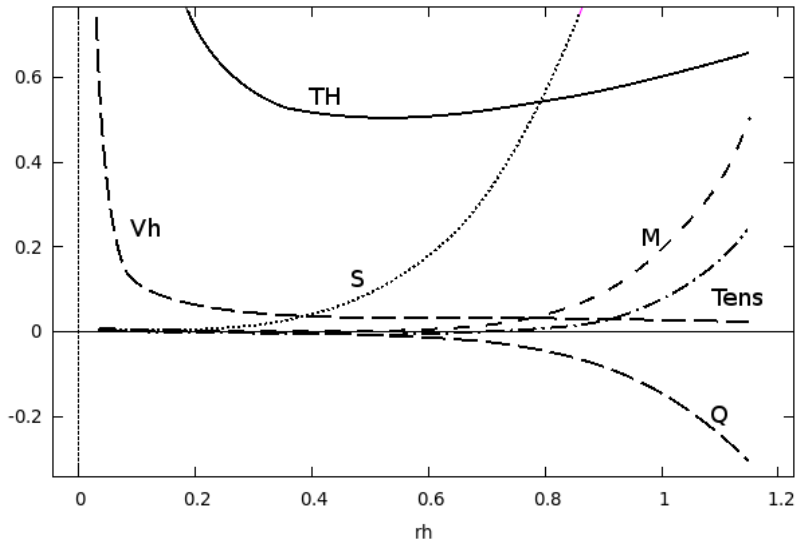}
\caption{Thermodynamical quantities of the charged $AdS$ branes for $n=2,\ p=3$ and $\ell=1$. The tension and angular momentum on the figure represent 
the total tensions $n\mathcal T$ and total angular momentum $(p+1)J/2$ respectively.}
\label{fig:thermon2p3_ch}
\end{figure}

We also present the various thermodynamical quantities in figure \ref{fig:thermon2p3_ch} for $\ell=1$ as a function of the horizon radius $r_h$. We note that the tension has a small negative region for small $r_h$.

\begin{figure}
 \centering
  \includegraphics[scale=.5]{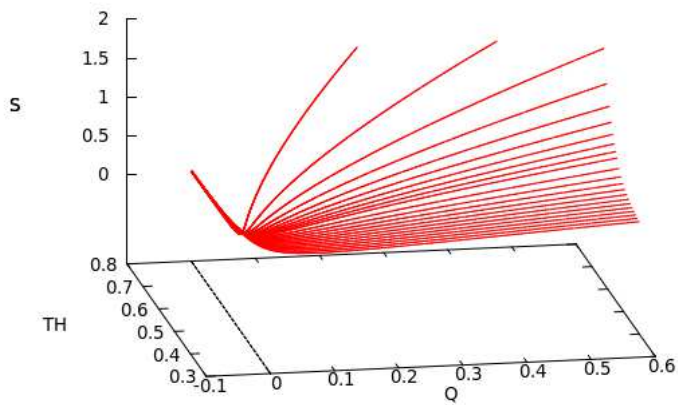}
\caption{The entropy as a function of the charge and the temperature for $p=3,n=2$.}
\label{fig:STHn2p3_ch}
\end{figure}

\begin{figure}
 \centering
  \includegraphics[scale=.5]{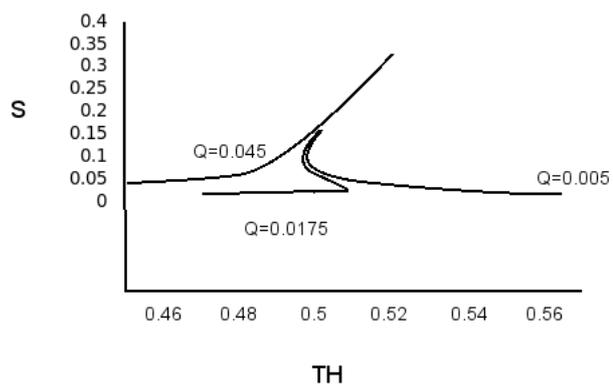}
\caption{Some constant $Q$ slices of the graphic presented in figure \ref{fig:STHn2p3_ch}.}
\label{fig:STHfQn2p3_ch}
\end{figure}

Althought Figure \ref{fig:thermon2p3_ch} seems to suggest the existence of thermally stable and unstable phases, it is not  fully generally the case: for fixed values of the charge, there seems to be only one thermally unstable branch for large values of the charge as illustrated on figure \ref{fig:STHn2p3_ch} where the entropy is plotted as a function of the charge and of the temperature. For small values of the charge however, it can be expected that thermally stable and thermally unstable branches exist. Note that this was already found in \cite{brs} for $n=1$. This is confirmed in figure \ref{fig:STHfQn2p3_ch} where the entropy as a function of the temperature is plotted for different fixed values of the charge.

\subsection{Rotating case}
Here only odd values of $p$ are allowed. We present our results for $\ell=1$. Our approach is the same as in the charged case, i.e. \emph{a posteriori} 
rescaling of the functions $a,b$ and consequently $w$. 

\begin{figure}
 \centering
  \includegraphics[scale=.5]{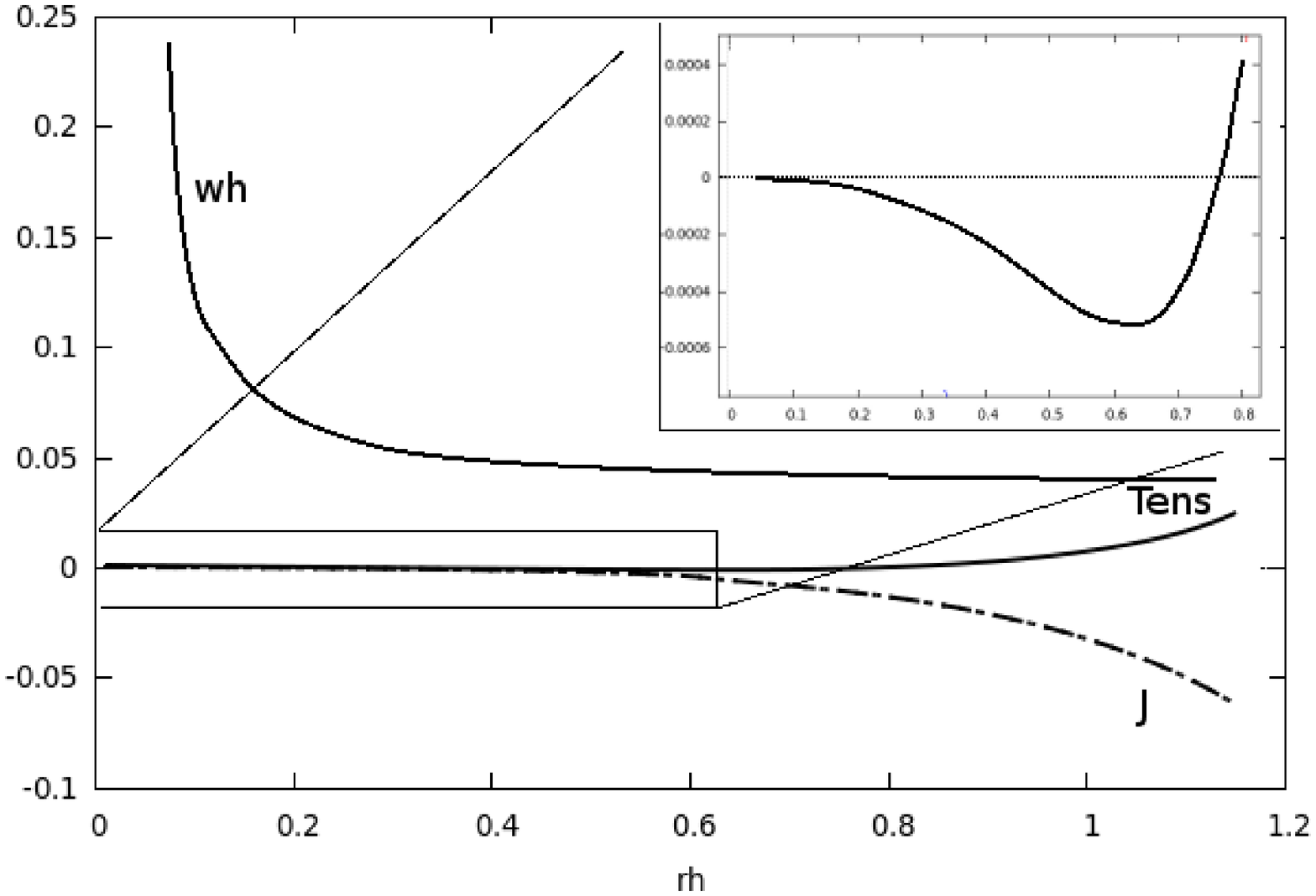}
\caption{A generic profile of the total tension, horizon angular velocity and total angular momentum as functions of the horizon radius. The box on upper right corner is a zoom on interval $[0,0.6]$. We systematically observed a region of negative tension. This was noticed and interpreted in \cite{adsnubs} in the uncharged and non rotating case for $n=1$.}
\label{fig:negtensn2p3}
\end{figure}

\begin{figure}
 \centering
  \includegraphics[scale=.5]{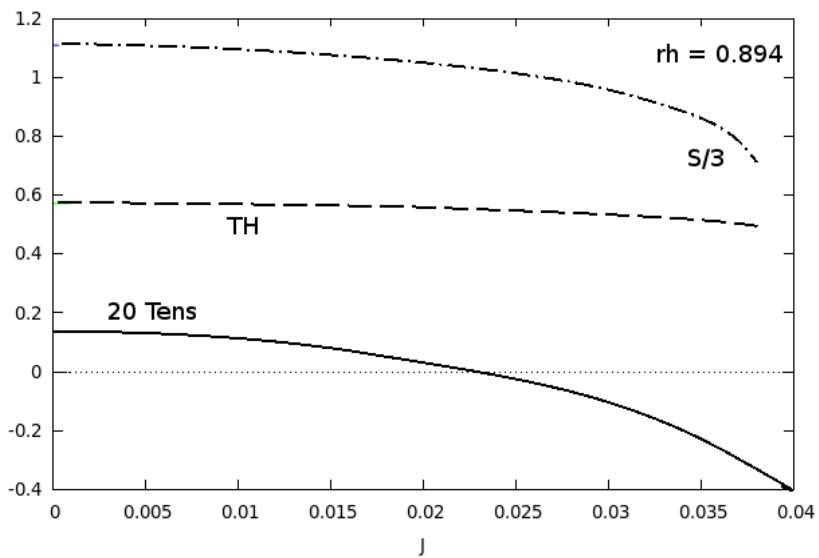}
\caption{The total tension decreases for increasing total angular momentum for fixed values of the horizon radius. Once the angular momentum is too high, the tension becomes negative. Note also that the entropy decreases with the angular momentum, as can be expected from the $AdS$ black holes. The mass is almost constant in the regime presented on the figure.}
\label{fig:tensn2p3}
\end{figure}

A striking feature of the rotating $AdS$ branes is that for small values of the angular velocity at the horizon ($w_h$), the tension is first negative, 
then changes sign and become positive. Above a critical value of the angular velocity (depending on the number of dimension), the tension is always negative. 
This can be interpretted as the balance between centrifugal force and gravitational tension; once the rotation is too high, the black brane wants to spread, 
this is an effect opposite to the tension. This is illustrated on figures \ref{fig:negtensn2p3} and \ref{fig:tensn2p3}.

\begin{figure}
 \centering
  \includegraphics[scale=.5]{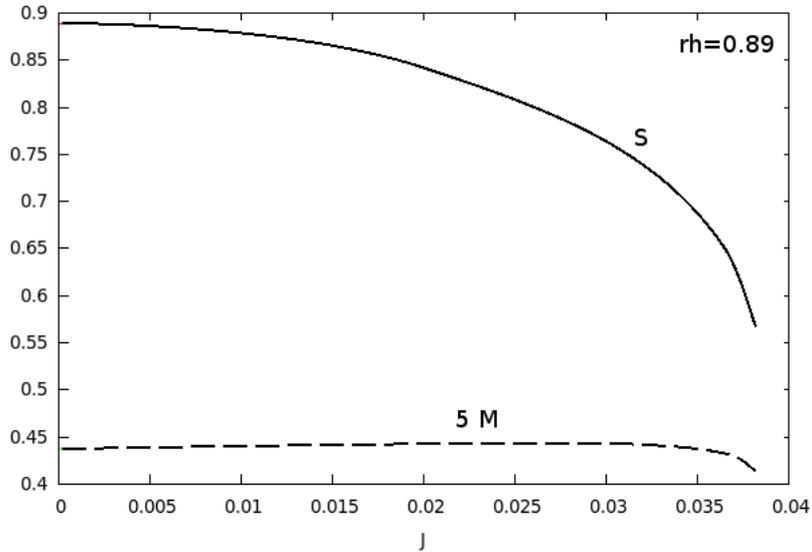}
\caption{The entropy as a function of the angular momentum for a fixed value of $r_h$. The mass is almost constant in the regime presented on the figure.}
\label{fig:sjn2p3}
\end{figure}

The entropy decreases with the angular momentum, similar to the case of rotating black holes. Again, this is to be understood as the spreading of the horizon 
with the rotation (at fixed mass). This is illistrated on figures \ref{fig:tensn2p3} and \ref{fig:sjn2p3}.

\begin{figure}
 \centering
  \includegraphics[scale=.5]{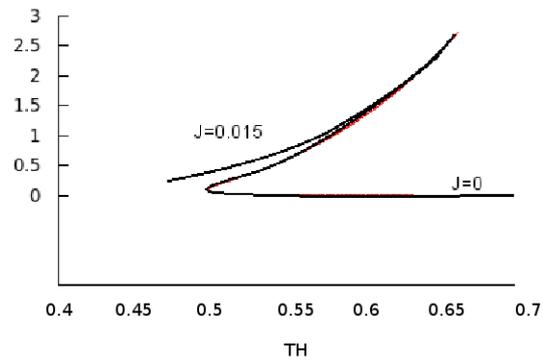}
\caption{Some constant $J$ slices of the graphic presented in figure \ref{fig:STHn2p3_ch}.}
\label{fig:STHfQn2p3_rot}
\end{figure}

The entropy as a function of the temperature for fixed values of the angular momentum behaves similar to the case of the rotating black strings \cite{brs}: 
there are two phases of $AdS$ black branes for vanishing values of the angular momentum and these phases quickly disappear for non vanishing values of $J$. We show the entropy as a function of the temperature and angular momentum in figure \ref{fig:STHfQn2p3_rot} and constant $J$ foliations of the latter in figure \ref{fig:STHfQn2p3_rot}.

\begin{figure}
 \centering
  \includegraphics[scale=.5]{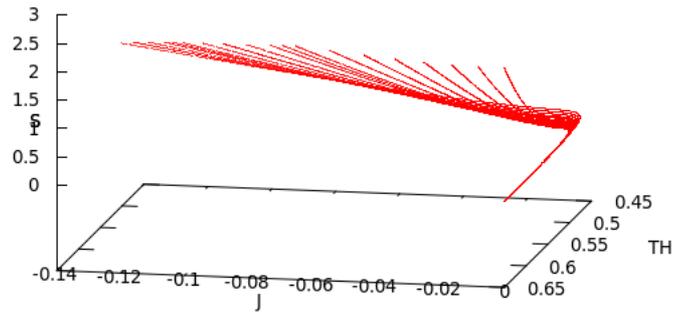}
\caption{The entropy as a function of the total angular momentum and the temperature for $p=3,n=2$.}
\label{fig:STHn2p3_rot}
\end{figure}

\subsection{Charged \emph{and} rotating case}
\begin{figure}
 \centering
  \includegraphics[scale=.5]{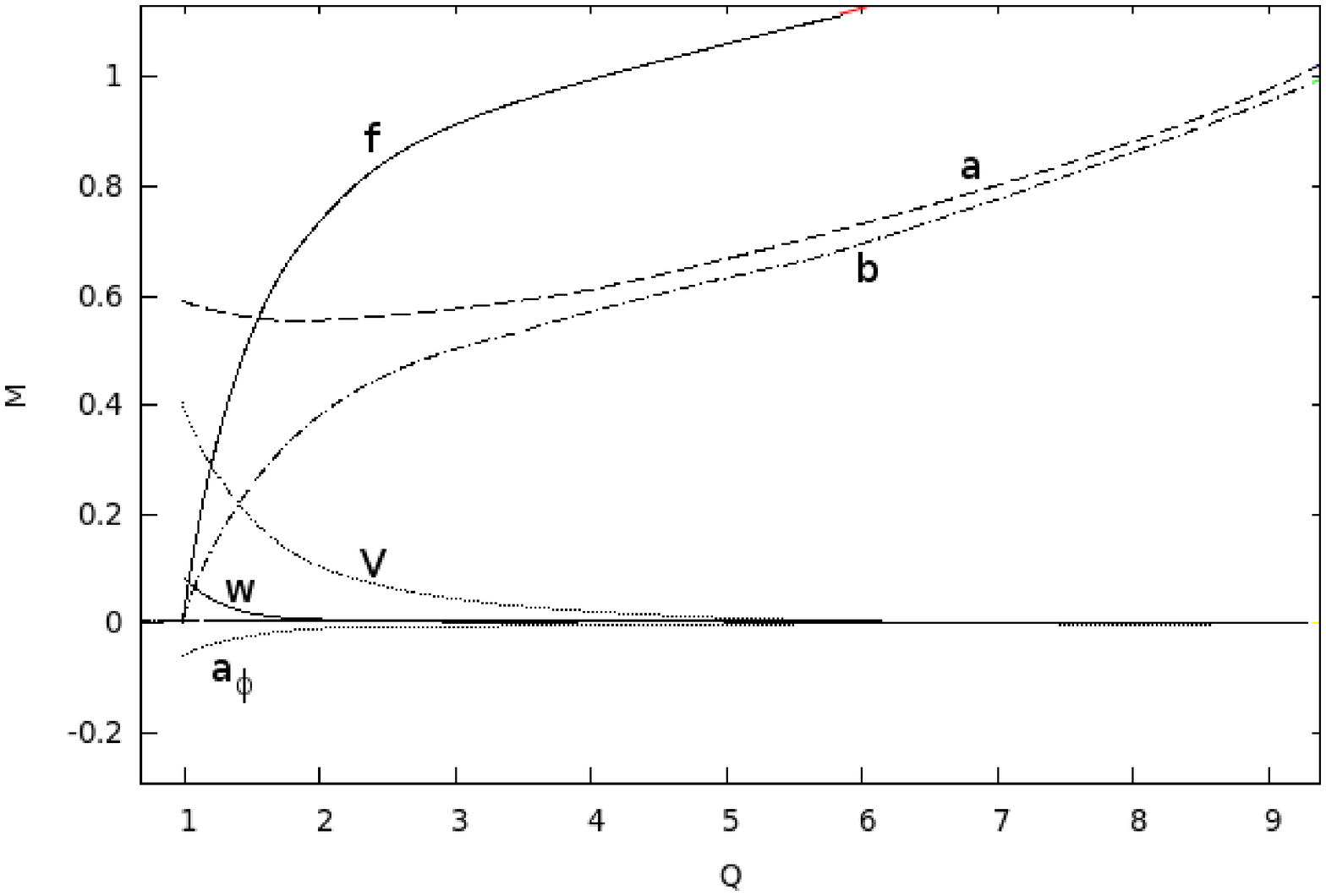}
\caption{A generic profile for the charged and rotating $AdS$ black brane with $n=2,\ p=3,\ \Lambda=-3$.}
\label{fig:profchrotn2p3}
\end{figure}

Finally we contructed the charged and rotating $AdS$ black brane solution, providing a numerical evidence that the solution indeed exists. The generic profile is presented in figure \ref{fig:profchrotn2p3} for $n=2,\ p=3,\ \Lambda=-3,\ r_h=1	$.
As discussed before, we don't discuss here its thermodynamical properties in details, due to the restrictions of our approach. The problem however deserves further investigations.

Let us finally mention that the tension is systematically negative for small values of the horizon radius as in the case $n=1$ \cite{adsnubs}, with or without charge and rotations.

\section{Conclusion}
In this paper, we presented evidences for the existence of a charged-rotating black brane in asymptotically locally $AdS$ spacetime. These solutions generalise the black branes considered in \cite{adsbrane}. The author of reference \cite{adsbrane} further considered the limit where the horizon radius shrinks to zero, namely the soliton limit and interpretted the resulting configuration in a braneworld context. Here we focused on the black brane solution but we believe that at least in the rotating case, the limit $r_h\rightarrow 0$ should exist. In the charged case, we have serious doubts about this since there exists a lower bound on the horizon radius (this was already discussed in \cite{brs} for $n=1$). Note that the $n$-dimensional manifold can be replaced by any $n$-dimensional Ricci flat manifold.

The lower bound on the horizon radius translates to a lower bound for the mass for fixed values of the charge. This is interpretted as the competition between the electric repulsion and the gravitational attraction. The tension has the same behaviour than the uncharged static case. In that sense, the charge influences the mass.

The tension is negative for small values of the horizon radius, this is to be understood in the same way as in ref \cite{adsnubs}: the spatial pressure of the cosmological constant dominates the gravitational tension due to the mass in the small $r_h$ regime while the situation is reversed for larger values of the horizon radius.

The rotation has the surprising effect of lowering the tension, even up to a point where the tension goes from positive to negative values. Here, the rotation seems to play a role opposite to the tension, even if the tension is concerned with the non rotating extradirections.

We did not study systematically the thermodynamics of the charged and rotating case, but we expect its feature to combine the properties of the charged case and of the rotating case.

Finally, let us mention that the charged case can be interpreted in the $AdS/CFT$ context, the value of $V$ at the origin being the source of a spin $1$ operator defined on a conformal field theory living in $\mathbb M^n\times S_p\times \mathbb R_t$ (recall that we can shift $V$ such that $V(r_h)=0$), the charge (next to leading order on the boundary) is then to be interpretted as the correlator of the spin $1$ field. In this context, the $S_p$ can be interpretted as an internal space and one the $\mathbb M^n$ can be thought as an infinite $n$-dimensional space. Such considerations are speculative but we believe these would deserve further investigations.

\section{Acknowledgement}
We gratefully acknowledge Y. Brihaye and E. Radu for useful discussion.


\begin{thebibliography}{90}
\bibitem{Maldacena:1997re}
J.~M.~Maldacena,
Adv.\ Theor.\ Math.\ Phys.\  {\bf 2} (1998) 231
[Int.\ J.\ Theor.\ Phys.\  {\bf 38} (1999) 1113]
[arXiv:hep-th/9711200].
\bibitem{Witten:1998qj}
E.~Witten,
Adv.\ Theor.\ Math.\ Phys.\  {\bf 2} (1998) 253
[arXiv:hep-th/9802150].
\bibitem{Hawking:1982dh}
S.~W.~Hawking and D.~N.~Page,
Commun.\ Math.\ Phys.\  {\bf 87} (1983) 577.

\bibitem{Witten:1998zw}
E.~Witten,
Adv.\ Theor.\ Math.\ Phys.\  {\bf 2} (1998) 505
[arXiv:hep-th/9803131].
\bibitem{Chamblin:1999tk}
A.~Chamblin, R.~Emparan, C.~V.~Johnson and R.~C.~Myers,
Phys.\ Rev.\ D {\bf 60} (1999) 064018
[arXiv:hep-th/9902170];
%
A.~Chamblin, R.~Emparan, C.~V.~Johnson and R.~C.~Myers,
Phys.\ Rev.\ D {\bf 60} (1999) 104026
[arXiv:hep-th/9904197].

\bibitem{Copsey:2006br}
  K.~Copsey and G.~T.~Horowitz, Gravity dual of gauge theory on S**2 x S**1 x R, JHEP {\bf 0606} (2006) 021 [arXiv:hep-th/0602003].
\bibitem{rms}
R. Mann, E. Radu and C. Stelea, Black string solutions with negative cosmological constant, JHEP09, 2006, 073
\bibitem{adsbrane}
  B.~Kleihaus, J.~Kunz and E.~Radu,New AdS solitons and brane worlds with compact extra-dimensions, arXiv:1006.3290 [hep-th].
\bibitem{Gibbons:2004uw}
G.~W.~Gibbons, H.~Lu, D.~N.~Page and C.~N.~Pope,Rotating black holes in higher dimensions with a cosmological constant, Phys.\ Rev.\ Lett.\  {\bf 93}, 171102 (2004) [arXiv:hep-th/0409155].
G.~W.~Gibbons, H.~Lu, D.~N.~Page and C.~N.~Pope,The general Kerr-de Sitter metrics in all dimensions,  J.\ Geom.\ Phys.\  {\bf 53}, 49 (2005) [arXiv:hep-th/0404008].
\bibitem{kunz}
J. Kunz, F. Navarro-Lerida, E. Radu, Higher dimensional rotating black holes in Einstein-Maxwell theory with negative cosmological constant, Phys. Lett. B649, 2007, 463-471
\bibitem{brs}
  Y.~Brihaye, E.~Radu and C.~Stelea, Black strings with negative cosmological constant: Inclusion of electric
  charge and rotation, Class.\ Quant.\ Grav.\  {\bf 24} (2007) 4839 [arXiv:hep-th/0703046].

\bibitem{mp}
R. Myers and M. Perry, Black holes In higher dimensional space-times, Annals Phys., 1986, 172, 304
\bibitem{bhchrot}
M. Cvetic, H. Lu, C. N. Pope, Charged Kerr-de Sitter black holes in five dimensions, Phys. Lett. B 273, 2004, 598

\bibitem{bhrev}
R. Emparan and H. S. Reall, Black holes in higher dimensions, Living Rev. Rel. 6, 2008, 11

\bibitem{gl}
R. Gregory and R. Laflamme, The instability of charged black strings and p-branes, Nucl. Phys. B428, 1994, 399-434
\bibitem{gubser}
S. Gubser, On non-uniform black branes, Class. Quant. Grav. 19, 2002, 4825-4844
\bibitem{wiseman}
T. Wiseman, Static axisymmetric vacuum solutions and non-uniform black strings, Class. Quant. Grav. 20, 2003
\bibitem{adsstab}
  Y.~Brihaye, T.~Delsate and E.~Radu, On the stability of AdS black strings, Phys.\ Lett.\  B {\bf 662} (2008) 264 [arXiv:0710.4034 [hep-th]]
\bibitem{bsrev}
T. Harmark, V. Niarchos and N. A. Obers, Instabilities of black strings and branes, Class. Quant. Grav. 24, 2007, R1-R90
\bibitem{counter}
    V.~Balasubramanian and P.~Kraus,A stress tensor for anti-de Sitter gravity, Commun.\ Math.\ Phys.\  {\bf 208} (1999) 413  [arXiv:hep-th/9902121].
\bibitem{potential}
M.~Cvetic and S. S. Gubser, J. High Energy Phys. {\bf 04}, 024 (1999); 
 M. M. Caldarelli, G.~Cognola and D. Klemm, Class. Quant. Grav. {\bf 17}, 399 (2000). 
\bibitem{colsys}
U. Ascher, J. Christiansen, R.~D. Russell, A collocation solver for mixed order systems of boundary value problems, Math. of Comp. {\bf 33} (1979) 659;  U. Ascher, J. Christiansen, R.~D. Russell, Collocation software for boundary-value ODEs, ACM Trans. {\bf 7} (1981) 209.
\bibitem{adsnubs}
  T.~Delsate,Non Uniform Black Strings and Critical Dimensions in $AdS_d$, JHEP {\bf 0907} (2009) 035, [arXiv:0904.2149 [hep-th]].



\end{thebibliography}
\end{document}